\documentclass{ecai} 

\usepackage{latexsym}
\usepackage{amssymb}
\usepackage{amsmath}
\usepackage{amsthm}
\usepackage{booktabs}
\usepackage{enumitem}
\usepackage{graphicx}
\usepackage{color}

\usepackage{subcaption}
\usepackage{multirow}
\newcommand{\parsection}[1]{\noindent\textbf{#1.}~}

\newcommand{\BibTeX}{B\kern-.05em{\sc i\kern-.025em b}\kern-.08em\TeX}

\begin{document}


\begin{frontmatter}


\paperid{123} 


\title{SmoothSinger: A Conditional Diffusion Model for Singing Voice Synthesis with Multi-Resolution Architecture}


\author[A]{\fnms{Kehan}~\snm{Sui}\thanks{Corresponding Author. Email: suikehan@gwu.edu.}}
\author[B]{\fnms{Jinxu}~\snm{Xiang}}
\author[C]{\fnms{Fang}~\snm{Jin}} 

\address[A]{George Washington University}
\address[C]{George Washington University}


\begin{abstract}
Singing voice synthesis (SVS) aims to generate expressive and high-quality vocals from musical scores, requiring precise modeling of pitch, duration, and articulation. While diffusion-based models have achieved remarkable success in image and video generation, their application to SVS remains challenging due to the complex acoustic and musical characteristics of singing, often resulting in artifacts that degrade naturalness. In this work, we propose SmoothSinger, a conditional diffusion model designed to synthesize high quality and natural singing voices. Unlike prior methods that depend on vocoders as a final stage and often introduce distortion, SmoothSinger refines low-quality synthesized audio directly in a unified framework, mitigating the degradation associated with two-stage pipelines. The model adopts a reference-guided dual-branch architecture, using low-quality audio from any baseline system as a reference to guide the denoising process, enabling more expressive and context-aware synthesis. Furthermore, it enhances the conventional U-Net with a parallel low-frequency upsampling path, allowing the model to better capture pitch contours and long term spectral dependencies. To improve alignment during training, we replace reference audio with degraded ground truth audio, addressing temporal mismatch between reference and target signals. Experiments on the Opencpop dataset, a large-scale Chinese singing corpus, demonstrate that SmoothSinger achieves state-of-the-art results in both objective and subjective evaluations. Extensive ablation studies confirm its effectiveness in reducing artifacts and improving the naturalness of synthesized voices.
\end{abstract}

\end{frontmatter}


\section{Introduction}
\label{sec:intro}

Deep learning has brought significant improvements to human audio generation, particularly in text-to-speech (TTS), where current state-of-the-art models produce speech that is vivid, natural, and almost indistinguishable from actual speech recordings. However, singing voice synthesis (SVS) continues to face considerable challenges. Unlike TTS, which handles relatively stable pitch and simple prosodic patterns, SVS must process complex musical structures with wider pitch ranges, frequent pitch transitions, and detailed timing requirements. 
Prior research in singing voice synthesis (SVS) has explored various components of the pipeline, including lyrics-to-singing alignment \cite{chien2016alignment, huang2022improving}, acoustic modeling \cite{lu2020xiaoicesing, nakamura2019singing, ren2020fastspeech}, adversarial synthesis \cite{chandna2019wgansing, hono2019singing}, and neural vocoding \cite{van2016wavenet, prenger2019waveglow}. More recently, the field has seen significant progress driven by the development of GAN-based and diffusion-based models \cite{kim2022adversarial, zhang2024stylesinger, zhang2024tcsinger}.

Recent progress in TTS and SVS has revealed two key challenges that hinder high-fidelity audio generation. The first relates to model architecture. Generative adversarial networks (GANs)~\cite{goodfellow2020generative} and denoising diffusion probabilistic models (DDPMs)~\cite{ho2020denoising} are among the most commonly used generative frameworks. While GANs have shown success in various tasks, their adversarial training often suffers from instability, where imbalances between the generator and discriminator can lead to convergence issues. In contrast, diffusion models offer a more stable alternative, gradually transforming Gaussian noise into structured data through a learned reverse process. Due to their robustness and strong generative capabilities, diffusion models have recently achieved state-of-the-art performance in both image and audio synthesis.

The second challenge lies in the design of acoustic–vocoder pipelines, particularly in diffusion-based SVS systems. Most existing methods adopt a two-stage framework in which an acoustic model first predicts a mel-spectrogram from musical score inputs (e.g., lyrics, pitch, duration), followed by a vocoder that reconstructs the waveform. While effective, this sequential structure introduces limitations:
\begin{itemize}
    \item The acoustic model and vocoder are typically trained independently. During inference, the vocoder receives generated mel-spectrograms, which may differ in distribution from the ground-truth spectrograms it was trained on, potentially degrading audio quality.
    \item Vocoder-based reconstruction introduces inherent artifacts such as background noise, distortion, or crackling. Even when using ground-truth spectrograms, the resulting waveform often fails to fully recover the original signal, suggesting that an end-to-end solution could better preserve fidelity.
\end{itemize}
These challenges highlight the need for unified, stable, and high-fidelity approaches to SVS that move beyond conventional two-stage paradigms.

In this paper, we propose SmoothSinger, a conditional diffusion model for high fidelity singing voice synthesis. Designed to produce smoother and more natural vocals with fewer artifacts, SmoothSinger introduces several key innovations to improve audio quality and model robustness:

\begin{itemize}

\item We incorporate a reference based dual branch architecture that leverages external audio as contextual guidance. In addition to the musical score and diffusion timestep \(t\), the model conditions on reference audio, which may come from any low quality model. This is implemented via a parallel branch initialized from the main U-Net, enabling context aware denoising and improved expressiveness.

\item We propose a Multi Resolution (MR) module that augments the standard U-Net with a parallel low frequency upsampling path. Unlike conventional sequential designs, this non sequential structure allows multi-scale features to directly influence the final output, enhancing spectral richness and reconstruction fidelity through efficient attention based processing.

\item To improve training stability, we replace the reference input with degraded ground truth audio during certain training steps. This mitigates temporal misalignment issues and encourages the model to generate cleaner outputs while preserving temporal coherence.

\end{itemize}


By incorporating these design principles, we develop an audio generation framework that removes the need for a separate vocoder, thereby reducing vocoder induced artifacts and addressing the mismatch issues common in two stage SVS pipelines. We evaluate SmoothSinger on the Opencpop dataset~\cite{wang2022opencpop}, a Chinese singing corpus, and show that it produces high quality synthesized singing voices. Noting that text-to-speech (TTS) is a simplified case of SVS with reduced conditioning, we further adapt the model to the TTS setting and evaluate it on the LJSpeech dataset~\cite{ljspeech17}. The generated speech closely aligns with the original recordings in both fidelity and naturalness, demonstrating the model’s versatility and effectiveness across tasks.


\section{Related Work}
\label{sec:relatedwork}

\subsection{Audio Processing}

Singing voice synthesis (SVS) has advanced significantly with the emergence of generative models based on adversarial and diffusion paradigms. Most existing SVS systems adopt a two-stage architecture in which an acoustic model predicts spectral representations from musical scores, followed by a vocoder that reconstructs the waveform. While this framework has shown success in models such as DiffSinger~\cite{liu2022diffsinger}, RDSinger~\cite{sui2024rdsinger}, and TCSinger~\cite{zhang2024tcsinger}, it often suffers from mismatches between stages, leading to artifacts and reduced expressiveness. Vocoders like HiFi-GAN~\cite{kong2020hifi} and DiffWave~\cite{kong2020diffwave} are widely used due to their efficiency and synthesis quality, but remain dependent on accurate alignment with acoustic features. Recent works such as FastDiff~\cite{huang2022fastdiff} and ProDiff~\cite{huang2022prodiff} explore unified architectures that directly map structured inputs to waveforms, reducing reliance on external vocoders. These approaches demonstrate improved temporal consistency and fidelity, and suggest a promising direction toward vocoder-free SVS.

\subsection{Singing Voice Synthesis}
Singing voice synthesis (SVS) primarily relies on two model architectures: GAN-based and diffusion-based models. GAN-based models, like VISinger2 \cite{zhang2022visinger2}, use adversarial training to produce smooth, harmonically consistent audio but often face training instability and mode collapse, limiting data diversity. Diffusion-based models, such as DiffSinger \cite{liu2022diffsinger}, provide fine-grained control over pitch and timing through iterative denoising, essential for capturing expressive transitions in singing. However, DiffSinger is not fully end-to-end, relying on a vocoder that may introduce alignment issues between generated features and the final audio, indicating room for quality improvement.



\subsection{Denoising Diffusion Probabilistic Models}

Denoising Diffusion Probabilistic Models (DDPMs)~\cite{ho2020denoising} have become a powerful class of generative models, offering stable training, strong sample diversity, and high quality outputs across image~\cite{dhariwal2021diffusion}, video~\cite{ho2022video}, and audio domains~\cite{kong2020diffwave,huang2022fastdiff}. Unlike GANs and VAEs, which often suffer from mode collapse or reduced output fidelity~\cite{goodfellow2020generative,kingma2019introduction}, DDPMs iteratively denoise data through a learned reverse process, allowing fine grained control and detailed generation. Conditional diffusion models extend this framework by incorporating auxiliary inputs, such as text~\cite{ramesh2022hierarchical}, acoustic features~\cite{huang2022fastdiff}, or structured signals like pose and depth maps~\cite{zhang2023adding}, enabling guided synthesis across complex modalities. More recent approaches improve controllability using reference-based conditioning, where aligned guidance inputs help enforce domain-specific structure. AnimateAnyone~\cite{hu2024animate} demonstrates precise motion synthesis by aligning reference images with diffusion stages, while RDSinger~\cite{sui2024rdsinger} applies a similar mechanism to singing voice synthesis by injecting reference audio into a two branch diffusion model. This alignment enhances expressive modeling of pitch and timing, and also accelerates convergence by sharing architectural components between the reference and generative paths.

\subsection{Multi-Resolution Processing}
The multi-resolution approach has become a cornerstone in generative modeling, allowing models to capture both fine-grained textures and broader structures, which are essential for producing realistic outputs. In image generation, recent advancements like DALL-E 2 \cite{ramesh2022hierarchical} employ multi-scale diffusion processes that progressively refine details at varying resolutions, improving fidelity and coherence. StyleGAN3 \cite{karras2021alias} addresses aliasing issues and maintains feature consistency across scales, resulting in enhanced image quality and smoother transitions between features. Furthermore, diffusion-based models such as SR3 \cite{saharia2022image} incorporate multi-resolution strategies specifically for super-resolution tasks, improving fine details in generated images by iteratively refining outputs at different scales. InstantNGP~\cite{muller2022instant} utilizes a multi-grid hash table construction, significantly improving both the computation efficiency and fidelity of 3D model reconstruction. Building upon this, Neural Impostor~\cite{liu2023neural} further develops a multi-resolution encoding algorithm on tetrahedral grids, enabling smooth motion of 3D models.

In audio synthesis, multi-resolution techniques have been equally impactful. Multi-band discriminators in HiFi-GAN \cite{kong2020hifi} and BigVGAN \cite{lee2022bigvgan} enable these models to capture both low-frequency structures, crucial for preserving tonal stability, and high-frequency details, enhancing the clarity of synthesized audio. Similarly, WaveGrad 2 \cite{chen2021wavegrad} extends this approach by introducing multi-scale processing that refines audio features across frequencies, providing greater realism and improved synthesis quality. These advancements highlight the adaptability and effectiveness of multi-resolution methods in achieving high-quality, coherent results across both visual and audio domains, emphasizing the role of multi-scale strategies in advancing generative tasks.


\section{Methods}
\label{method}

\begin{figure*}
  \centering
  \begin{subfigure}{0.30\linewidth}
    \includegraphics[width=\hsize]{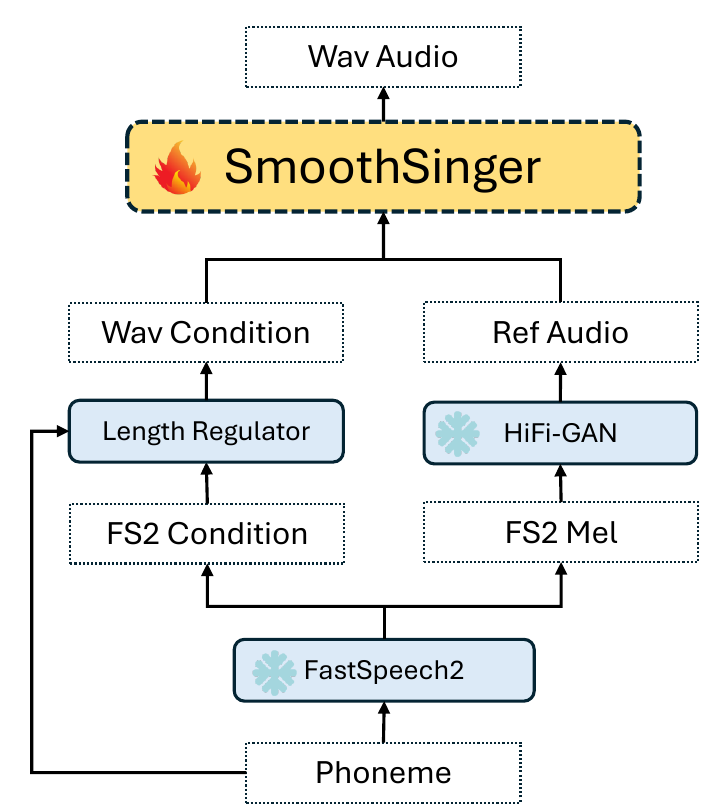}
    \caption{The overall process flow of generation.}
    \label{fig:short-a}
  \end{subfigure}
  \hfill
  \begin{subfigure}{0.68\linewidth}
    \includegraphics[width=\hsize]{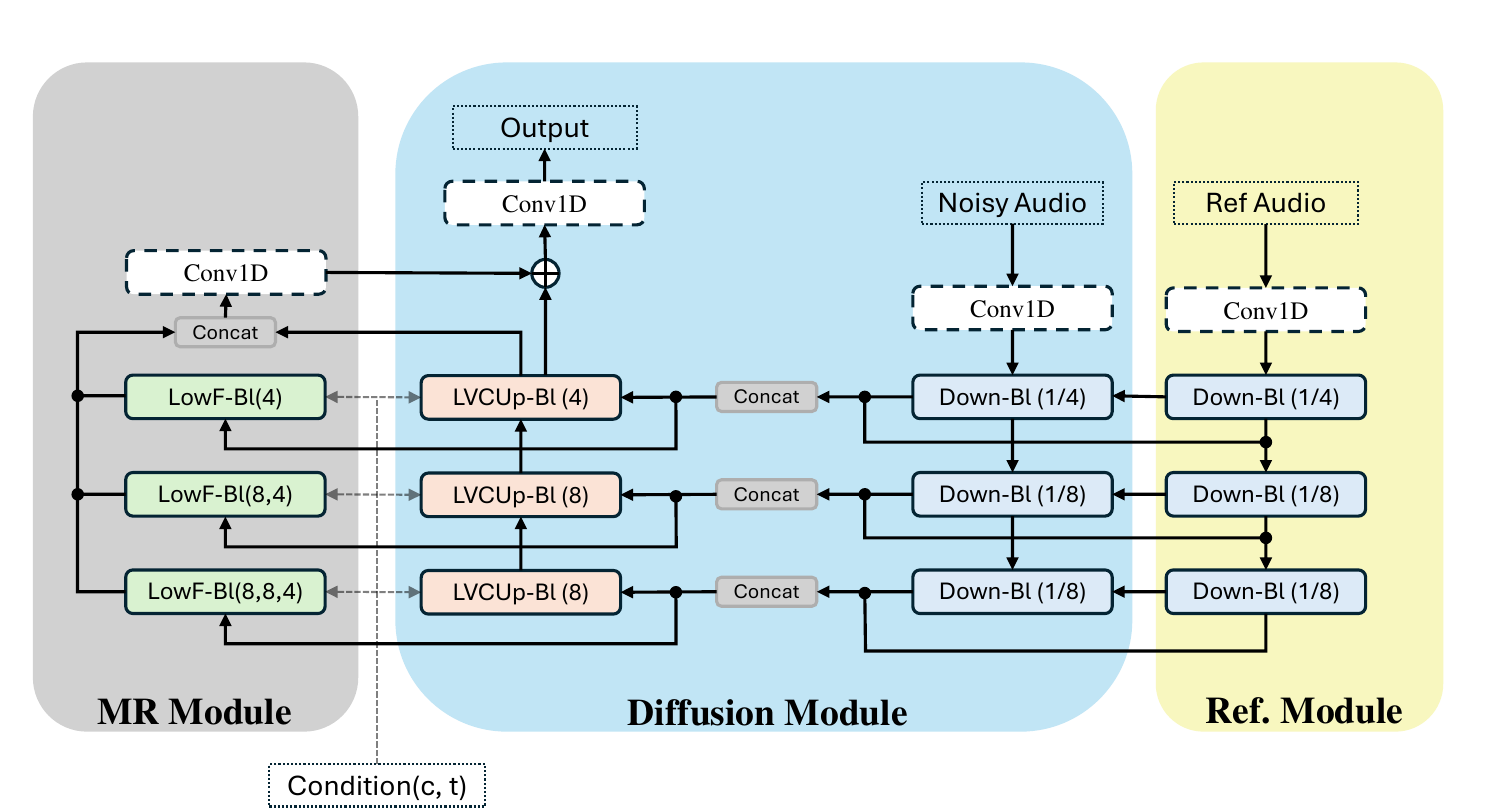}
    \caption{The main architecture of SmoothSinger consists of three modules: the MR module, the Diffusion module, and the Reference module.}
    \label{fig:short-b}
  \end{subfigure}
  \vspace{3ex}
  \caption{(a) illustrates the complete process of audio generation from phoneme input, with the internal structure of SmoothSinger shown on the right. The Length Regulator adjusts features of varying lengths by repeating them to match the length of the waveform. In (b), "Down-Bl" represents the downsampling blocks used in our model, which process both noisy and reference audio inputs using the same structure. "LVCUP-Bl" refers to the time-aware location-variable convolution upsampling block, following the architecture detailed in FastDiff~\cite{huang2022fastdiff}. "LowF-Bl" denotes the multi-resolution block for handling low-frequency features, with further details provided in Figure~\ref{fig:LowF}.}
  \label{fig:model}
  \vspace{4ex}

\end{figure*}

This work focuses on singing voice synthesis, and in this section, we present the complete design of the proposed model, SmoothSinger. The overall workflow is illustrated in Figure~\ref{fig:short-a}. As an initial step, the input conditions such as lyrics, musical score, and speaker ID are first processed by a baseline model. We adopt FastSpeech2 due to its popularity and strong performance in previous studies, while alternative choices for the baseline model are listed in Table~\ref{tab:exp:ablref}. The main architecture of SmoothSinger, shown in Figure~\ref{fig:short-b}, consists of three core modules: the Diffusion module described in Section~\ref{sec:diffusionmodule}, the Reference module detailed in Section~\ref{sec:referencemodule}, and the MR module explained in Section~\ref{sec:mrmodule}. The use of degraded data for training is further described in Section~\ref{method:degraded}.

\subsection{Preliminariy: Diffusion Model}
\label{sec:diffusionmodule}
Diffusion models \cite{ho2020denoising} are probabilistic generative models that sequentially transform a simple noise distribution into a complex data distribution through a series of gradual transformations. The process typically involves two stages: the forward process and the reverse process.

\subsubsection{Forward Process}
The forward process, or diffusion process, gradually adds noise to the input data over multiple time steps. Let \(x_0\) be the original sample, and \(x_t\) denote the data at time step \(t\) with accumulated noise. This process forms a Markov chain with transitions:

\begin{equation}
    q(x_t | x_{t-1}) = \mathcal{N}(x_t; \sqrt{1 - \beta_t} x_{t-1}, \beta_t I)
\end{equation}

where \(\beta_t\) is a noise schedule controlling the Gaussian noise added at each step. As \( t \) approaches the final step \( T \), the data converges to an isotropic Gaussian distribution.

\subsubsection{Reverse Process}
The reverse process, also called the denoising process, learns to recover the data distribution by iteratively denoising from \( x_T \) back to \( x_0 \). The reverse process approximates the true posterior \( q(x_{t-1} | x_t) \) through a parameterized model \( p_\theta(x_{t-1} | x_t) \), which is typically implemented as a neural network with parameters \( \theta \). The reverse process can be expressed as:
\begin{equation}
    p_\theta(x_{t-1} | x_t) = \mathcal{N}(x_{t-1}; \mu_\theta(x_t, t), \Sigma_\theta(x_t, t)),
\end{equation}
where \( \mu_\theta \) and \( \Sigma_\theta \) represent the mean and variance of the Gaussian distribution learned by the model. During training, the model minimizes a variational bound on the negative log-likelihood, which leads to the following loss function:
\begin{equation}
    L_\text{vlb} = \mathbb{E}_{q(x_{1:T}|x_0)} \left[ \sum_{t=1}^T D_{KL}(q(x_{t-1}|x_t, x_0) \, || \, p_\theta(x_{t-1} | x_t)) \right]
\end{equation}
where \( D_{KL} \) is the Kullback-Leibler divergence.

\begin{figure}
  \centering
  \includegraphics[width=0.9\hsize]{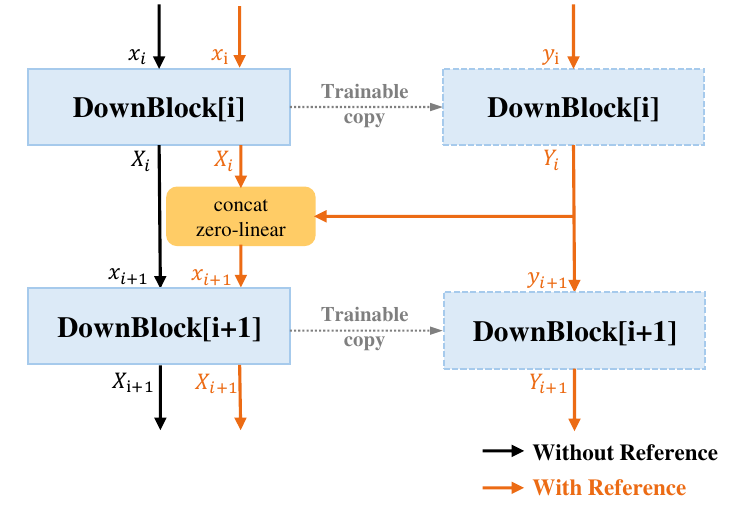}
  \caption{The structure of the Reference module is illustrated, with the black arrow indicating the data flow without reference guidance and the yellow arrow showing the flow when the Reference module is incorporated.}
  \label{fig:ref}
  \vspace{3ex}
\end{figure}

\subsection{Reference Based Diffusion}
\label{sec:referencemodule}
In singing voice synthesis (SVS), the input to the model typically consists of music score embeddings, which may include phonetic content, pitch, note durations, and other musical descriptors. While this representation captures the symbolic structure of the target audio, it lacks direct acoustic grounding. In contrast, the mel-spectrogram serves as a more explicit and perceptually aligned representation for waveform reconstruction, as it directly bridges the gap between symbolic features and the vocoder’s input space. To enrich the model's contextual awareness and facilitate more expressive synthesis, we introduce a reference-based feature extraction mechanism into the diffusion architecture. 

As illustrated in Figure~\ref{fig:ref}, the left branch presents two downsampling blocks, offering a simplified representation of the downsampling pathway in the diffusion module. Following initial training with the original diffusion model, the architecture is duplicated to form a reference branch on the right. All parameters in this auxiliary branch are initialized with the weights from the original model, and the weight update is conducted independently for each branch. The reference audio is fed into the downsampling blocks of the auxiliary branch to extract multi-resolution features. These features are then transferred to the main branch within the same resolution and passed into the subsequent downsampling block. 

Let the subscript \( i \) indicate the \( i \)-th block from the top in the model structure shown in Figure~\ref{fig:short-b}. Denote \( x_i \) and \( X_i \) as the input and output of the \( i \)-th downsampling block in the Diffusion module, and \( y_i \) and \( Y_i \) as the corresponding input and output in the Reference module. Prior to incorporating reference guidance, the model follows a sequential propagation where $x_{i+1} = X_i$. After integrating the reference pathway, the propagation is modified as:
\begin{equation}
\label{eq1}
    x_{i+1} = f(\text{concat}(X_i, Y_i)),
\end{equation}
where $f(\cdot)$ is implemented as a zero-initialized $1 \times 1$ convolution layer that learns a weighted fusion of the main and reference features.

The upsampling trajectory consists of a sequence of time-aware LVC blocks, as introduced in FastDiff~\cite{huang2022fastdiff}. Let \( Z_i \) denote the output of the \( i \)-th LVCUp block from the top. Each LVCUp block takes the output from the next lower-resolution block \( Z_{i+1} \), the concatenated features from the downsampling path \( f(\text{concat}(X_i, Y_i)) \), and the embedded condition \(\text{Condition}(c, t)\). The computation can be expressed as:
\begin{equation}
Z_i = g(Z_{i+1}, f(\text{concat}(X_i, Y_i)), \text{Condition}(c, t)),
\end{equation}
where \( g(\cdot) \) denotes the transformation applied by the LVCUp block, and \( f(\cdot) \) is defined as in Equation~\ref{eq1}.

This dual-branch structure enables precise retention and reuse of high-level acoustic information from reference signals, which is progressively decoded through the upsampling path. The result is an enhanced generative process capable of producing high-fidelity, expressive singing voices with improved coherence across temporal and stylistic dimensions. A comparable design is ControlNet~\cite{zhang2023adding}. However, a key distinction lies in the integration strategy: while ControlNet injects conditioning only into the upsampling trajectory, our approach feeds the output of the reference module into both the downsampling and upsampling paths of the diffusion model. Our design enables the model to incorporate a richer set of features from the reference audio, leading to improved reconstruction quality. The effectiveness of this approach is demonstrated in the performance results shown in Table~\ref{tab:exp:ablcomp}.


\subsection{Multi-Resolution Low-Frequency Architecture}
\label{sec:mrmodule}

Most diffusion-based generative models adopt the U-Net architecture, characterized by symmetric downsampling and upsampling paths. This structure inherently captures multi-resolution information and expands the model's receptive field. Unlike conventional U-Net designs, the core architecture of SmoothSinger extends the standard framework by incorporating an additional low-frequency upsampling module, referred to as the MR module in Figure~\ref{fig:short-b}, which enables more effective capture and integration of spectral patterns across diverse frequency bands.

Let \( K_i \) denote the output of the \( i \)-th low-frequency block from the top. The output of each block in the MR module is defined as:
\[
K_i = \Theta(f(\text{concat}(X_i, Y_i)),\ \text{Condition}(c, t)),
\]
where \( \Theta(\cdot) \) represents the transformation applied within a LowF-Block, and \( K_i \in \mathbb{R}^L \) for any \( i \), with \( L \) denoting the dimensionality of the audio signal at the original resolution.

In the standard diffusion U-Net architecture, upsampling blocks are sequentially connected, and each stage predicts finer-resolution features from the previous coarser level. In contrast, the MR module adopts a non-sequential design, wherein low-frequency blocks are independently connected to the final output. This structure enables multi-scale inputs to directly influence the high-resolution reconstruction, thereby enriching the synthesized signal with complementary spectral information from varying temporal contexts. The effectiveness of the MR module is evaluated in Table~\ref{tab:exp:ablcomp} of the ablation study.

\begin{figure}
  \centering
  \includegraphics[width=0.5\hsize]{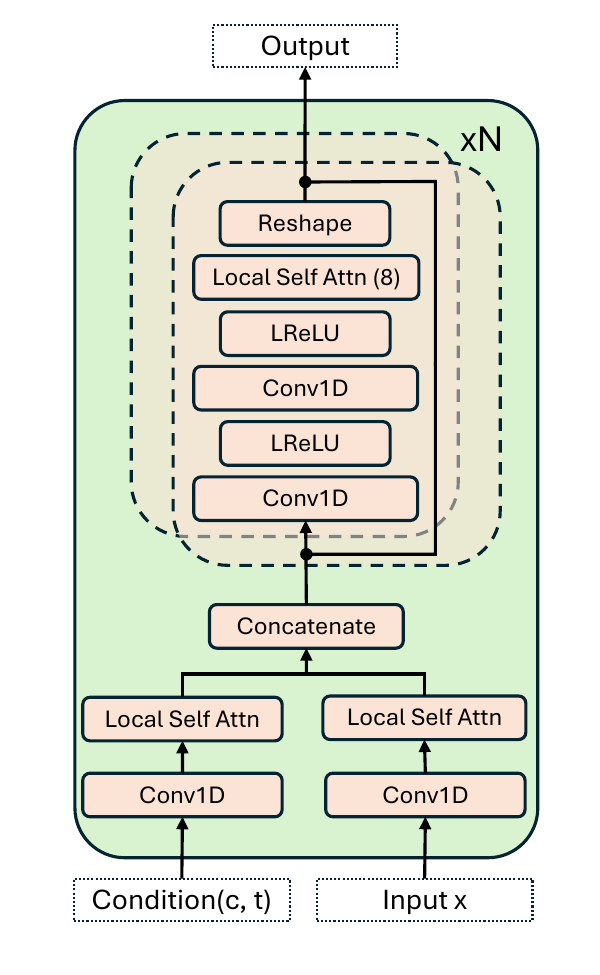}
  \caption{The network structure of the Low-Frequency Upsampling Block. The local self-attention employs a sliding window attention mechanism, as outlined in Longformer~\cite{beltagy2020longformer}. The notation "(8)" indicates that the hidden features are upscaled by a factor of 8, achieving the upsampling effect after reshaping.}
  \label{fig:LowF}
  \vspace{4ex}
\end{figure}

Figure~\ref{fig:LowF} provides detailed information about the construction of the low-frequency upsampling block. This module uses 1D convolution and self-attention to process the audio hidden states. Given that waveforms can be long, using attention with quadratic complexity is impractical. Therefore, we apply a sliding window approach to restrict the attention mechanism to local information, achieving a computational complexity of $O(L)$, where $L$ is the length of audio. The low-frequency block performs upsampling by increasing the hidden size and reshaping. We believe this approach more effectively preserves low-frequency information within this branch, while other audio characteristics, such as transitions between frequency ranges and pitch variations, are modeled by the LVC upsampling block.

\begin{table*}[h]
\centering
\begin{tabular}{l|cccccc}
\toprule
Method & MOS (↑) & Ranking (↓) & SIG MOS (↑) & BAK MOS (↑) & PESQ (↑) & STOI (↑) \\
\midrule
GT & 4.71 ± 0.04 & - & 4.07 ± 0.04 & 4.38 ± 0.04 & - & - \\
GT.recon & 4.35 ± 0.04 & - & 3.76 ± 0.04 & 4.29 ± 0.04 & 3.72 ± 0.05 & 0.953 ± 0.009 \\
\midrule
FastSpeech2 & 3.02 ± 0.07 & 3.98 ± 0.12 & 3.19 ± 0.05 & 3.93 ± 0.07 & 2.94 ± 0.07 & 0.882 ± 0.018 \\
Diffsinger & 3.26 ± 0.06 & 3.33 ± 0.11 & 3.33 ± 0.06 & 4.05 ± 0.06 & 3.09 ± 0.06 & 0.907 ± 0.013 \\
RDSinger & 3.52 ± 0.07 & 2.56 ± 0.10 & 3.41 ± 0.05 & \textbf{4.12 ± 0.07} & 3.16 ± 0.07 & 0.932 ± 0.016 \\
VISinger2 & 3.43 ± 0.06 & 2.77 ± 0.09 & 3.45 ± 0.06 & 3.84 ± 0.06 & \textbf{3.31 ± 0.07} & 0.925 ± 0.011 \\
SmoothSinger & \textbf{3.61 ± 0.06} & \textbf{2.36 ± 0.10} & \textbf{3.47 ± 0.06} & 4.08 ± 0.06 & 3.25 ± 0.07 & \textbf{0.937 ± 0.013} \\
\bottomrule
\end{tabular}
\caption{Objective and Subjective Evaluation Metrics for SVS Models}
\label{tab:exp:main}
\vspace{4ex}
\end{table*}

\subsection{Enhanced Training with Degraded Audio}
\label{method:degraded}

In generative models, synthetic data is widely used to augment training due to its abundance, accessibility, and variability, which help improve generation quality. SmoothSinger uses reference audio to guide synthesis, where mel-spectrograms are generated and then converted to audio. However, this introduces temporal misalignment between ground-truth audio and the reference audio, making loss computation unstable. To mitigate this, we replace the reference audio with degraded versions of the ground-truth audio during certain training steps, enabling the diffusion model to learn temporal alignment more effectively.

The degradation methods include adding noise, adjusting amplitude, and distorting high or low frequencies. We randomly select several regions within the training waveform, where both the number of regions and their lengths are determined randomly. The number of the region is uniformly sampled from $3$ to $10$ and the length of each region is sampled from a uniform range between $500$ and $2000$ units. For each selected region, we apply a random combination of the following degraded methods.

\parsection{Add Random Noise}
Let \(X\) be the clean audio signal. We perturb it with Gaussian noise \(\mathcal{N}(0,\sigma^2)\) scaled by \(\alpha\), yielding
\[
X_{\mathrm{noisy}} = X + \alpha\,\mathcal{N}(0,\sigma^2).
\]
In our experiments, we fix \(\sigma=1\) and sample  
\[
\alpha \sim \text{Uniform}(0.8,\,1.05)
\]

\parsection{Adjust amplitude}
The amplitude of the audio signal is scaled by a random factor \(\beta\). This transformation can be described mathematically as follows:

\begin{equation}
X_{adject}(t) = \begin{cases} 
\beta \cdot X(t), & \text{if } t \in \text{selected region}, \\
X(t), & \text{otherwise},
\end{cases}
\end{equation}

where \( X(t) \) is the original audio signal, \( X_{adject}(t) \) is the transformed signal, \( \beta \) is the scaling factor sampled from \( \beta \sim \text{Uniform}(0.8, 1.05) \), and \( t \) denotes the time index of the audio signal.

\parsection{Apply distortion} 
The amplitude of the audio signal is transformed by applying a random exponential factor. The use of an exponential factor allows subtle adjustments to the dynamic characteristics of the audio signal, simulating natural variations in loudness or timbre while preserving the signal's integrity.

\begin{equation}
X_{\text{distort}}(t) = 
\begin{cases} 
\text{sign}(X(t)) \cdot |X(t)|^{\gamma}, & \text{if } t \in \text{selected region}, \\
X(t), & \text{otherwise},
\end{cases}
\end{equation}

where \( \gamma \) is the exponential factor sampled from \( \gamma \sim \text{Uniform}(0.9, 1.2) \),

\parsection{Change strength in different frequency} 
The amplitude of randomly selected frequency bands in the audio signal is modified using the Short-Time Fourier Transform (STFT) and inverse STFT (ISTFT). To ensure consistent audio length, the selection of the audio region is determined based on the STFT parameters. This guarantees that the length of the audio after applying the ISTFT matches the original length. The process is defined as follows:

First we apply the STFT to the audio signal \( X(t) \), resulting in the frequency-domain representation \( X_{\text{STFT}}(f, \tau) \):
\begin{equation}
X_{\text{STFT}}(f, \tau) = \text{STFT}(X(t)),
\end{equation}
where \( f \) represents the frequency bins, \( \tau \) represents the time frames, and the STFT is computed with a hop size of 128, a frame size of 512, and a Hann window function.

Then we randomly select a set of frequency bands \( \mathcal{F} = \{f_i, f_j, \dots \} \), and scale their amplitudes:
\begin{equation}
X_{\text{scaled}}(f, \tau) = 
\begin{cases} 
\sigma(f) \cdot X_{\text{STFT}}(f, \tau), & \text{if } f \in \mathcal{F}, \\
X_{\text{STFT}}(f, \tau), & \text{otherwise},
\end{cases}
\end{equation}
where \( \sigma(f) \) is a random scaling factor for each frequency band, sampled from \( \sigma(f) \sim \text{Uniform}(0.8, 1.1) \).

Finally, we apply the ISTFT to transform the modified frequency-domain representation back to the time domain:
\begin{equation}
X_{\text{ISTFT}}(t) = \text{ISTFT}(X_{\text{scaled}}(f, \tau)),
\end{equation}
using the same parameters as in the STFT.


\section{Experiments}
\label{experiment}

This section begins with a description of the experimental setup, followed by the presentation and analysis of the main results for SmoothSinger. An ablation study is then conducted to further evaluate the contributions of each component to overall model performance.

\subsection{Experimental Setup}
\subsubsection{Data Processing}
The Opencpop dataset~\cite{wang2022opencpop} is a publicly available, high-quality Mandarin singing corpus crafted for applications in singing voice synthesis (SVS). This dataset comprises 100 unique songs performed by a professional female vocalist, amounting to approximately 5.2 hours of audio content. Although it includes only a single singer, we assign a speaker ID of 0 and embed it as a conditioning signal, following the approach in DiffSinger. The recordings were made in a studio environment to ensure high fidelity. Each recording is accompanied by detailed phonetic annotations, including syllables, note pitch, duration, and phonemes. The authors have also released their official textgrid with audio segmentation. In this segmentation, the dataset is divided into 3,756 utterances, each ranging from 5 to 20 seconds in duration. Opencpop selects 5 out of 100 songs as the test set, comprising 206 utterances, while the remaining 95 songs with 3,550 utterances are used for training. We follow the same train-test split as defined by Opencpop to ensure consistency.

In our pipeline, the target ground truth is represented as a waveform with a 24 kHz sampling rate. Our framework utilizes FastSpeech2~\cite{ren2020fastspeech} and HiFi-GAN~\cite{kong2020hifi} to generate reference audio, as FastSpeech2 serves as a baseline model in numerous audio processing frameworks, including DiffSinger and FastDiff, while HiFi-GAN is the most widely used vocoder in these models. The intermediate mel-spectrogram generated by FastSpeech2 has a hop size of 128 samples, a frame size of 512 samples, and an 80-dimensional mel-spectrogram. HiFi-GAN then converts this mel-spectrogram back to our reference audio using the same parameters. Additionally, we use Parselmouth~\cite{parselmouth} to extract the ground-truth $F_0$ (fundamental frequency)~\cite{wu2020adversarially}, and use the PyPinyin library~\cite{ren2020deepsinger} to decompose Chinese lyrics into individual syllables.

\subsubsection{Model Configuration}
SmoothSinger consists of four main components as shown in Figure~\ref{fig:short-b}. Two down-sampling modules are used to handle diffusion noise and reference audio, and two up-sampling modules, the LVC upsampling module and the low-frequency upsampling module, process audio features across different frequency bands. Our model utilizes the same up-sampling and down-sampling rates of [8,8,4], with a total of 16M trainable parameters.

\subsubsection{Training Setup}
\label{sec:train_setup}
Our pipeline consists of two main steps. First, we generate reference audio using FastSpeech2 and HiFi-GAN, during which no variables require training. In the second step, SmoothSinger is trained for 1.6M steps until convergence. During the first 1.2M steps, training is conducted on standard reference audio, while in the final 400K steps, degraded ground truth audio is involved with a $50\%$ probability. The audio used in training has a sample rate of 24 KHz, with lengths randomly selected between 25.6K and 51.2K, all in multiples of 256 to ensure correct down-sampling. We use the AdamW optimizer with $\beta_1 = 0.9$, $\beta_2 = 0.98$, and a learning rate that gradually decreases from $10^{-4}$ to $10^{-6}$. Training was conducted on a single NVIDIA 4090 GPU with a batch size of 16 over approximately 72 hours.


\subsection{Main Results}
\subsubsection{Evaluation Metrics}
\label{experiment:metrics}
To evaluate the performance of different models, we used several quantitative metrics. PESQ~\cite{rix2001perceptual} and STOI \cite{taal2011algorithm} are widely used in audio processing, especially for assessing the quality and intelligibility of speech signals. Although waveform alignment can pose a challenge in SVS tasks, minor temporal discrepancies do not affect the reliability of these metrics, which remain effective in reflecting audio quality trend. Additionally, we employed the DNSMOS~\cite{reddy2022dnsmos} metric, an objective and non-intrusive measure of speech quality that uses deep learning to predict subjective scores as given by human listeners. Specifically, SIG MOS and BAK MOS are used in our experiments to evaluate speech quality and background noise, respectively. 

In addition, we employed human evaluation methods that commonly used in speech synthesis assessment. We used the MOS (Mean Opinion Score) to evaluate model performance, with a scoring scale from 1 to 5. A total of 16 participants were invited to listen to synthesized audio samples and rate them within this range. Additionally, following the evaluation approach in RDSinger~\cite{sui2024rdsinger}, we conducted a blind ranking test. For each audio segment, participants were presented with outputs from different models and asked to rank them based on audio quality, where a rank of 1 indicates the audio with best quality. This metric is labeled as 'Ranking' in all tables. While MOS is included as a metric, the addition of Ranking is motivated by the fact that MOS scores often yield similar values across models for high-quality audio. Ranking, however, requires participants to select the preferred audio, providing a clearer indication of preference among models. Each metric is presented with its mean score and $95\%$ confidence interval.

\subsubsection{Analysis}
We present a comparative evaluation of SmoothSinger's audio quality against that of competing models, including FastSpeech2~\cite{ren2020fastspeech}, DiffSinger~\cite{liu2022diffsinger}, RDSinger~\cite{sui2024rdsinger}, and VISinger2~\cite{zhang2022visinger2}. Each of these models relies on a vocoder to convert generated mel-spectrograms into audio. For consistency, we uniformly apply HiFi-GAN as the vocoder in our comparisons, which is also the neural vocoder used in each of these original studies. To reconstruct ground truth, we use the same mel-spectrogram and HiFi-GAN combination. To ensure reliable and fair comparisons, we provide the same audio conditions as input across all models.

Table~\ref{tab:exp:main} presents the primary results of our comparative experiments. In this table, GT.recon refers to the reconstruction of ground-truth audio, serving as an upper bound for all acoustic-vocoder models. Our proposed model, SmoothSinger, achieves superior performance in both subjective metrics, with MOS and ranking scores exceeding those of competing models by 0.09 and 0.20, respectively, indicating a higher perceptual audio quality. SmoothSinger also leads in SIG MOS and STOI metrics, and achieves on-par performance with state-of-the-art models across other metrics. 


\subsection{Ablation Studies}
\label{sec:ablation}

\begin{table}[h]
\centering
\begin{tabular}{l|ccc}
\toprule
Method & MOS & BAK MOS \\
\midrule
Full Model* & 3.61 ± 0.06 & 4.08 ± 0.06 \\
W/o Ref. DownBlock & 3.31 ± 0.07 & 3.98 ± 0.07 \\
W/o LowF. UpBlock & 3.49 ± 0.06 & 4.05 ± 0.06 \\
W/o Degraded Data & 3.16 ± 0.07 & 3.72 ± 0.08 \\
\bottomrule
\end{tabular}
\caption{Ablation study on the different components}
\label{tab:exp:ablcomp}
\end{table}

\begin{table*}[h]
\centering
\begin{tabular}{l|ccccc}
\toprule
Method & MOS (↑) & SIG MOS (↑) & BAK MOS (↑) & PESQ (↑) & STOI (↑) \\
\midrule
GT & 4.81 ± 0.03 & 4.19 ± 0.03 & 4.50 ± 0.03 & - & - \\
GT.recon & 4.62 ± 0.03 & 3.96 ± 0.04 & 4.32 ± 0.03 & 3.91 ± 0.05 & 0.973 ± 0.007 \\
\midrule
FastSpeech2 & 3.67 ± 0.06 & 3.44 ± 0.06 & 4.01 ± 0.07 & 3.35 ± 0.06 & 0.925 ± 0.011 \\
Grad-TTS & 3.83 ± 0.05 & 3.50 ± 0.06 & \textbf{4.14 ± 0.06} & 3.56 ± 0.06 & \textbf{0.954 ± 0.010} \\
FastDiff & \textbf{3.94 ± 0.05} & 3.52 ± 0.05 & 4.09 ± 0.06 & \textbf{3.59 ± 0.06} & 0.948 ± 0.009 \\
SmoothSinger & 3.88 ± 0.06 & \textbf{3.57 ± 0.05} & 4.13 ± 0.06 & 3.52 ± 0.06 & 0.947 ± 0.009  \\
\bottomrule
\end{tabular}

\caption{Objective and Subjective Evaluation Metrics for TTS Models}
\label{tab:exp:tts}
\end{table*}


\begin{table}[h]
\centering
\begin{tabular}{l|cccc}
\toprule
Strides & SIG MOS & BAK MOS & PESQ \\
\midrule
$[8, 8, 4]$* & 3.47 ± 0.06 & 4.08 ± 0.06 & 3.25 ± 0.07 \\
$[8, 8, 8]$ & 3.41 ± 0.05 & 4.04 ± 0.06 & 3.17 ± 0.06 \\
$[4, 4, 4, 4]$ & 3.45 ± 0.06 & 4.08 ± 0.06 & 3.27 ± 0.07 \\
$[16, 16]$& 3.24 ± 0.06 & 3.92 ± 0.06 & 3.01 ± 0.07 \\
\bottomrule
\end{tabular}
\caption{Ablation study on the different downsampling stride}
\label{tab:exp:ablstride}
\end{table}

\begin{table}[h]
\centering
\begin{tabular}{l|cccc}
\toprule
Ref. Model & SIG MOS & BAK MOS & PESQ \\
\midrule
FastSpeech2* & 3.47 ± 0.06 & 4.08 ± 0.06 & 3.25 ± 0.07 \\
DiffSinger & 3.42 ± 0.06 & 4.13 ± 0.06 & 3.26 ± 0.07 \\
RDSinger & 3.49 ± 0.06 & 4.15 ± 0.06 & 3.20 ± 0.07 \\
\bottomrule
\end{tabular}
\caption{Ablation study on the different reference audio input}
\label{tab:exp:ablref}
\end{table}

\begin{table}[h]
\centering
\begin{tabular}{l|c|cc}
\toprule
Method &  Steps & SIG MOS & BAK MOS \\
\hline
\multirow{3}{*}{DiffSinger} & 4 & 3.14 ± 0.08 & 3.82 ± 0.07 \\
                            & 24 & 3.33 ± 0.06 & 4.05 ± 0.06 \\
                            & 100 & 3.37 ± 0.06 & 4.13 ± 0.06 \\
\hline
\multirow{3}{*}{RDSinger} & 4 & 3.21 ± 0.07 & 3.94 ± 0.07 \\
                          & 24 & 3.41 ± 0.05 & 4.12 ± 0.07 \\
                          & 100 & 3.44 ± 0.05 & 4.10 ± 0.07 \\
\hline
\multirow{3}{*}{SmoothSinger} & 4 & 3.29 ± 0.07 & 3.91 ± 0.07 \\
                              & 24* & 3.47 ± 0.06 & 4.08 ± 0.06 \\
                              & 100 & 3.48 ± 0.06 & 4.14 ± 0.06 \\
\bottomrule
\end{tabular}
\caption{Comparison of Denoising Steps of DiffSinger, RDSinger and SmoothSinger}
\label{tab:exp:steps}
\end{table}

We conducted ablation studies to demonstrate the effectiveness of various design choices in SmoothSinger, with the proposed model marked with an asterisk (*) at the end. Our model structure is relatively complex, so we first analyze the contributions of each module to verify their importance within the model. To do this, we conducted ablations by removing the reference module from the downsampling stage and the multi-resolution low-frequency processing module from the upsampling stage. In these configurations, all interactions between the model and these modules were replaced with zero-value inputs, and the model was retrained accordingly. Additionally, we examined the impact of removing the technique of training with degraded ground-truth audio, while keeping the overall model structure intact. The results in Table~\ref{tab:exp:ablcomp} illustrate the significance of these modules. Due to the temporal misalignment between the reference audio and the ground truth, the absence of degraded data aligned with the ground-truth audio leads to a substantial drop in performance.

Directly using waveforms can lead to overly long one-dimensional audio sequences, making it difficult for convolutional layers to capture comprehensive information. Our model adopts a UNet-based down-up sampling structure, where we employ a three-layer downsampling stride of [8,8,4] to compress the audio sequence length. Table~\ref{tab:exp:ablstride} presents the results of training with different stride configurations. Most models achieve comparable performance with these settings, indicating that our current choice is effective.


Our model takes the outputs generated by FastSpeech2 and HiFi-GAN as reference audio inputs, positioning SmoothSinger as an enhancement model for existing generative audio models. Both DiffSinger and RDSinger report improved performance over FastSpeech2; therefore, we use their outputs as alternative reference audio sources. However, as shown in Table~\ref{tab:exp:ablref}, using these references in SmoothSinger does not lead to a notable improvement in audio quality. As DiffSinger and RDSinger are both built based on the result of FastSpeech2, they result in significantly slower computation speeds compared to FastSpeech2 itself. These findings suggest that FastSpeech2, with its faster processing speed, remains a more efficient and practical choice for generating reference audio in this context.


In diffusion models, the selection of denoising steps significantly impacts the quality of audio generation. Like DiffSinger and RDSinger, SmoothSinger is also based on a diffusion framework. We define the denoising steps integer at 4 steps by following the noise scheduling and sampling approach in FastDiff~\cite{huang2022fastdiff}, and generate denoising step integers for 24 and 100 steps using a linear schedule. Table~\ref{tab:exp:steps} presents the performance of these models across different denoising steps, showing that SmoothSinger consistently achieves superior generation quality at each experiment.


\subsection{Text-to-Speech Synthesis}
\label{application}

To demonstrate the robustness of the proposed model, we also applied it to the Text-to-Speech Synthesis (TTS) task. Unlike SVS, since TTS has fewer audio conditions, we adjusted the number of input channels in the first layer of the conditioning structure’s convolution, keeping all other parameters unchanged. Starting from the pretrained SmoothSinger model, we fine-tune with a learning rate of $10^{-5}$ for 24 hours to obtain our TTS model.

We use the LJSpeech dataset~\cite{ljspeech17} to conduct comparisons on the TTS task, evaluating several models: FastSpeech2~\cite{ren2020fastspeech}, Grad-TTS~\cite{popov2021grad}, and FastDiff~\cite{huang2022fastdiff}. To ensure consistency, we use identical conditioning inputs across these networks. Unlike in the SVS task, the TTS task does not include pitch conditions, so we set these inputs to zero in our model. Table~\ref{tab:exp:tts} shows that our results achieve a comparable level to state-of-the-art models across multiple metrics in TTS task.




\section{Conclusion}
\label{conclusion}


Due to the two-stage design in most existing singing voice synthesis (SVS) models, the generated audio often contains noticeable artifacts. To address this limitation, we propose SmoothSinger, an end-to-end SVS model built upon the diffusion probabilistic framework. The model introduces a novel dual downsampling and dual upsampling architecture, extending the conventional diffusion model, which is essentially a U-Net. The additional downsampling branch, referred to as the Reference module, leverages reference audio to extract acoustic features at multiple resolutions and enhances control during the reconstruction process. The additional upsampling branch, denoted as the MR module, adopts a non-sequential design that independently gathers features from each resolution and integrates them into the final prediction. To further address the prevalent misalignment issue in SVS, we incorporate degraded audio during training, which significantly improves the alignment between intermediate features and the final waveform. Experimental results demonstrate that SmoothSinger achieves superior performance in both objective and subjective evaluations compared to state-of-the-art models in SVS and TTS tasks, producing audio with higher fidelity, naturalness, and overall quality.




\bibliography{mybibfile}

\end{document}